# Geographic Information Systems in Evaluation and Visualization of Construction Schedule

V. K. Bansal and Mahesh Pal
Department of Civil Engineering, National Institute of Technology,
Kurukshetra, Haryana, India -136119
vijaybansal18@yahoo.com

**Abstract:** Commercially available scheduling tools such as *Primavera* and *Microsoft Project* fail to provide information pertaining to the spatial aspects of construction project. A methodology using geographical information systems (GIS) is developed to represent spatial aspects of the construction progress graphically by synchronizing it with construction schedule. The spatial aspects are depicted by 3D model developed in *AutoCAD* and construction schedule is generated using *Microsoft Excel*. Spatial and scheduling information are linked together into the GIS environment (*ArcGIS*). The GIS-based system developed in this study may help in better understanding the schedule along with its spatial aspects.

**1.0 Introduction**

The bar charts and networks are widely used to depict the construction schedule whereas different activities in the schedule are linked with one or more components of the project under consideration. Bar charts/ networks provide the non-spatial information that lacks in spatial aspects of the different construction activities. Thus, to have the spatial aspects of a project, the construction planner uses 2D drawings and associates its different components with the related activities present in the schedule (Koo and Fischer 2000). Further, there is no dynamic linkage between a schedule and its spatial aspect in the commercially available scheduling tools such as *Primavera* and *Microsoft Project*. Interpretation of the schedule without any link of its activities with the corresponding spatial components is cumbersome as an actual project may contain thousands of activities, which makes difficult to check the schedule sequence for its completeness. Thus generating a gap in effective communication among different project participants. This limitation of schedule forced the researchers to combine scheduling tools with 3D CAD systems to depict the construction sequence visually in 3D that leads to the development of the 4D CAD. Koo and Fischer (2000) suggested that a 4D model increases the comprehensibility of the project schedule, thus, allowing users to detect mistakes or potential problems prior to the construction. Despite a lot of research in 4D CAD technology, their use is not very common in the construction industry as these tools are somewhat difficult to use and cannot be manipulated by everyone (Issa et al. 2003).

Within last decade, GIS tools with inbuilt 3D display have become more widely accessible to mainstream practitioners. Several researc h suggest the usefulness of GIS in construction industry to effectively handle various construction project requirements such as data management, integrating information, visualization, cost estimate, site layout and construction planning etc. (Bansal and Pal 2006 a,b,c; Cheng and O'Connor 1996; Varghese and O'Connor 1995; Zhong et al. 2004). GIS is found to be helpful in improving the construction planning and design efficiency by integrating the spatial and non -spatial information in a single environment (Jeljeli et al. 1993; Camp and Brown 1993; Oloufa et al. 1994). Recently, Poku and Arditi (2006) supplemented the non-spatial scheduling techniques by developing a GIS based spatial system to represent construction progress that is synchronized with construction schedule. They also discussed the important issue of construction scheduling and progress control that concerns with practice of construction management. They used design generated in *AutoCAD* and plugged with schedule (generated using *primavera* ) into a GIS package. The study by Poku and Arditi (2006) sheds light on how to supplement the project management softwares, concluded that integrating GIS with project management tools such as *Primavera* require backend coding to make the existing system more user friendly. Present study explores the potential of GIS in evaluation and visualization of construction schedule as well as to maintain the construction data within GIS environment that can be associated with the activities in the schedule.

**2. GIS-Based Operations**

The *AutoCAD* is used to generate the spatial data (2D data layers) corresponding to each activity in the schedule. In addition *ArcGIS* also contain the tools to handle the editing session (ArcGIS 2004). The functionalities/features of *ArcGIS* used in this study are discussed below.

## 2.1 Spatial Operation

The *merge* in *ArcGIS* is used to group the features of a layer into one feature. It combines different features by removing boundaries or nodes between adjacent polygons. The non -adjacent polygons in same layer are also merged to create a multipart polygon feature. The *Base height* is the elevation value for the features of a layer in 3D space. The 2D layer does not have *base height* and *feature height* information. To display 3D perspective view, features of a 2D layer generated in *AutoCAD* are assigned *base height* and *feature height* from the fields of its own *attribute table*. The *extrusion* tool in *ArcGIS* is also used to changes the points into vertical lines, lines into vertical walls, and polygons in to 3D blocks. *ArcGIS* is also used to create a new layer by piecing together two or more layers of the same geometry (ArcGIS 9 2004).

## 2.2 Maintaining and Integration of Construction Resources Data in GIS

*ArcGIS* is used to maintain the construction resources data in tabular form and integrate this data with the corresponding activities of the project. This approach also replaces the manual methods to extract the information from the database and allows easy updating, as most of the information is in digital form (Bansal and Pal 2006 b, c). Three types of relationships defined in the *ArcGIS one-to-one*, *one-to-many*, and *many-to-one* are used in this work to join different tables together (Chang 2002). The *Join* function of *ArcGIS* is used to establish a *one-to-one* or *many-to-one* relationship between the destination table and the source table (ArcGIS 9 2006). Two tables are joined on the basis of a field called *Activity_ID* that is available in both the tables. The name of the field does not have to be the same in both tables, but the data type must be same (allows joining number-to-number or string-to-string). The function *Relate* establishes *one-to-many* relationship between the destination table and the source table.

## 3.0 Procedure for the Evaluation and Visualization of Schedule

The detailed procedure to construct spatial aspects in *AutoCAD* corresponding to each activity in schedule is discussed in more detail by Bansal and Pal (2006c). The procedure to evaluate and visualize the construction schedule is discussed below:

***Step 1:*** *Generation and transfer of construction schedule-*schedule is a model of how and when the various tasks of a project are going to be accomplished. The schedule acts as a roadmap for the successful implementation of a project (Moder et al.1983). Typically, a project is divided into discrete activities and the time needed to complete each activity is decided and are arranged in sequential or overlapping order. The *Microsoft Excel* is used as scheduling tool (Hegazy and Ayed 1999), to determine the starting /finishing times, different floats, project duration and the critical activities. The schedule developed in*Microsoft Excel* is then transferred into the *ArcGIS*.

***Step 2:*** *Generation and transfer of 3D components* - spatial information of different activities is generated in *AutoCAD,* which form the basis of the proposed system. The *ArcGIS* allows working with the drawings generated in *AutoCAD*. The drawings are transferred to *ArcGIS* as layers and can be symbolized and queried. *ArcGIS* uses the *AutoCAD* drawing layers like any other type of feature layer. However, to edit or modify a CAD drawing layer's features or its associated attribute table, layers have to be converted to *shapefiles*. The s*hapefiles* are simple, non-topological format for the storing geometric location and attribute information of geographic features.

***Step 3:*** *Merging components/features* -components transferred into *ArcGIS* from *AutoCAD* may be merged together according to the activities as defined earlier in schedule generated in *Microsoft Excel*. Thus, the components of the drawing that belong to the same activity but are located at different positions in the space are joined together to construct the spatial data for each activity.

***Step 4:*** *Connecting of the schedule with corresponding 3D components-* this step involves in adding a field called *Activity_ID* to the schedule table and the *attribute* table of each component. The field *Activity_ID* is common between two tables (i.e. schedule table and *attribute* table of different components) and used to establish the connection between the component and the corresponding activity in the schedule. All the entries in the field *Activity_ID* are to be entered manually and should be unique in both schedule and *attribute* tables of an activity. Thus, the attribute required to associate the components with the corresponding activities in the schedule are the entries in the field *Activity_ID*.

*Step 5: Schedule Evaluation and corrections* -This step involves in evaluating the schedule to see the order of construction sequence. The schedule for different dates can be evaluated by visualizing it in the 3D space. If it complies with the actual construction sequence desired and does not requires any change in number of activities and logic, the schedule will finally be accepted and no alteration is allowed afterward. If the model does not comply with the required construction sequence and need some changes in the logic, the schedule is changed again as per the requirements in the *Microsoft Excel*. After the implementation of desired changes in the construction schedule, it is again associated with the related components for its evaluation. The sequence of the construction schedule is again checked to verify its sequence. Further, if it does not require any change, it is finally accepted. Sometime the different steps need to be repeated again if the numbers of activities in the schedule need addition/deletion as well as if some changes are required in the corresponding components generated in the *AutoCAD*.

## 4.0 Advantages of Scheduling in GIS Environment

Planner makes schedule in such a way that all components of the project must have related activities. By viewing the schedule, it becomes quite difficult to determine whether the schedule is complete. However, to confirm that all components of the project have related activity is a time consuming process because of the large number of activities in the network. In GIS a visual check of schedule is possible, which may help in preventing omission of the activities in the schedule. With non-spatial schedule it becomes difficult to predict if the activities are within/out of the construction sequence, because of the activities with mutual dependencies (i.e., successor and predecessor relationships) may be located in different parts of the schedule. Such non-spatial schedules force users to visualize and interpret the activity sequence in their minds. Therefore, multiple participants of project must individually conceptualize the sequence by associating the activities with the components shown in drawings. The schedule in GIS displays at 'what time' and 'where' in the space the components are to be built.

GIS facilitate the understanding of 3D model and topological relationship between different components in many ways (like zooming, pan, fly forward or backward, navigation etc). Any components can be set transparent that makes it easier to visualize the model. The users also have the option of rotating 3D components around the $x$, $y$, or z-axes to observe the developed 3-D models. Further, the element can also be viewed from any direction and angle. Integrating information such as project schedules and drawings allows visual understanding of the construction process, and the construction data can be linked with the corresponding activities. The failure or success of a building contract largely depends on the quality and timing of the information available to the contractors from the database. Thus, require an automated system to get the required information without delay. Present methodology links the construction resources data of various activities with the corresponding activities (Bansal and Pal 2006b). The proposed methodology also supports the cost estimation that involves quantity takeoffs and the cost of various resources required in the construction project. Some of the existing approaches for quantity takeoffs are not accurate enough to eliminate the possibility of errors like missing or duplicating different items of work. GIS-Based cost estimation methodology eliminates the errors like missing or duplicating various items by visualizing each components corresponding to the items in 2D or 3D space. Thus, it may help in increasing the productivity of quantity estimator by reducing the manual work in determining the quantity takeoffs. Accurate bill of quantities can also be generated on the basis of dimensions of various data layers within GIS (Cheng and Yang 2001). A detailed methodology of GIS based cost estimation is proposed by Bansal and Pal (2006 c). From schedule, the quarries like activities starting on the particular date and activities starting between the particular intervals of time can be made in the proposed model. The project activities can be easily stored and listed in variety of useful ways, such as sorting schedule in ascending or descending order on any field (floats, early or late start time) in the table. Selected records could be displayed in same table by promoting them to the top. The system still requires automation in some tasks, such as transferring information from *Microsoft Excel* and *AutoCAD* to *ArcGIS*, in current version of system these operations are performed manually.

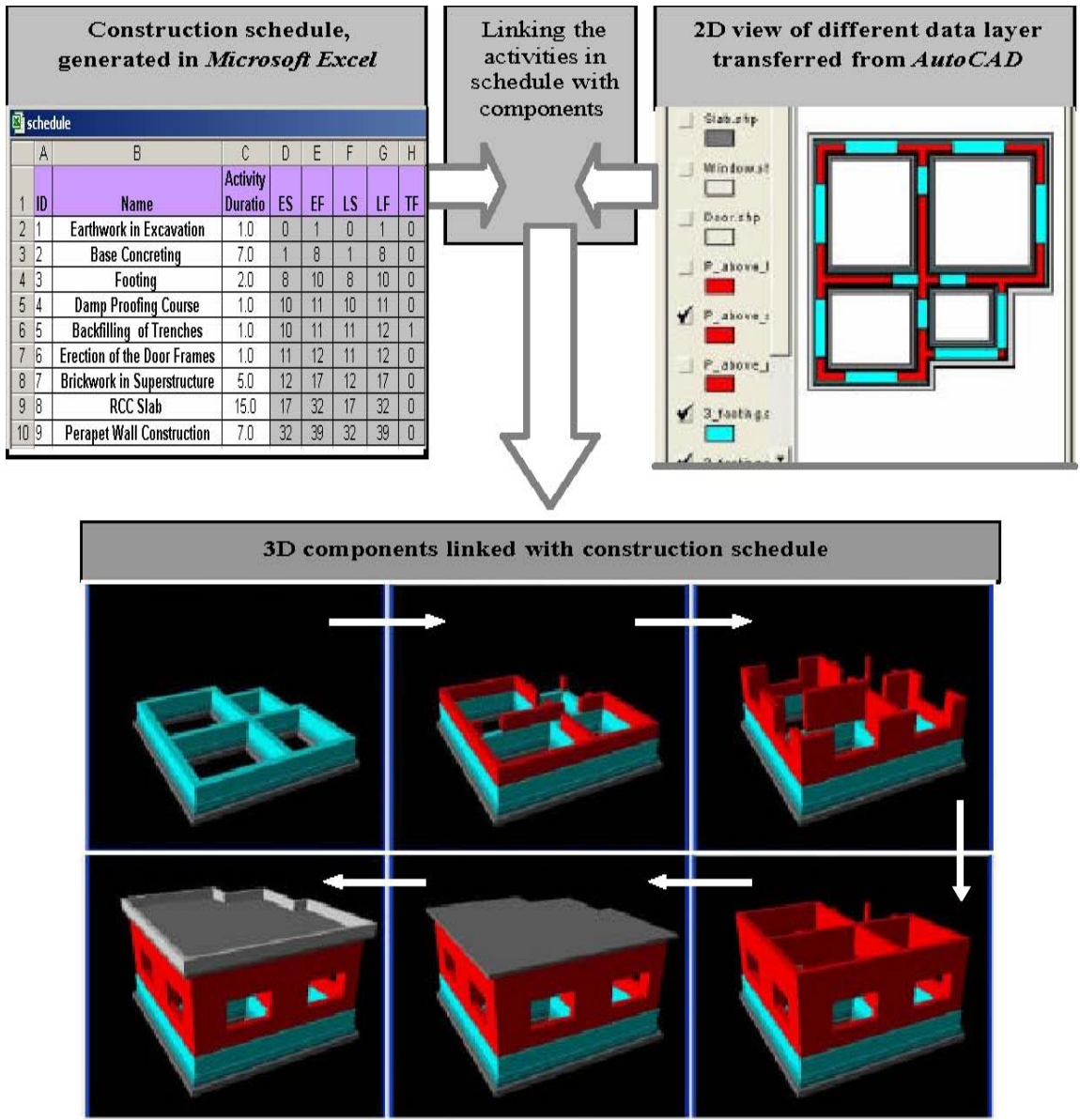

Figure 1: Linking the 3D model with construction schedule

**5.0 Conclusions**

This paper purposes a methodology to use GIS in representing and integrating the spatial and non-spatial information on the single environment. Methodology integrates construction schedule with corresponding spatial details so as to make understanding of the project sequence easier. The link allows easier understanding of the project as well as helps to detect possible problems in it. The proposed methodology supports additional analyses like rate analysis, cost estimates, and allows integrating safety recommendation with critical activity, thus ma king schedules more realistic (Bansal and Pal 2006a,b). Non-spatial schedules can only convey what is built 'when', whereas the schedule in GIS conveys what is being built 'when and where'. Thus, proposed work concludes that GIS can effectively be used for construction scheduling evaluation also.